\documentclass[preprint,aps,amssymb]{revtex4} 
\usepackage{graphicx}%
\newcommand{\unbfm}[1]{\mbox{\bf #1}}

\begin{document}

\title{Bayesian analysis of magnetic island dynamics}

\author{R.\ PREUSS}
\email{preuss@ipp.mpg.de}
\author{M.\ MARASCHEK}
\author{H.\ ZOHM}
\author{V.\ DOSE}
\affiliation{Max-Planck-Institut f\"ur Plasmaphysik, EURATOM Association\\
Boltzmannstr.\ 2, D-85748 Garching b. M\"unchen, Germany}

\date{\today}

\begin{abstract}
We examine a first order differential equation with respect to time coming
up in the description of magnetic islands in magnetically confined plasmas.
The free parameters of this equation are obtained by employing Bayesian
probability theory.
Additionally a typical Bayesian change point is solved in the process of
obtaining the data.
\end{abstract}

\maketitle

\section{Introduction}

Magnetic islands are structures appearing on resonant surfaces of
plasmas in toroidal magnetic confinement devices.
They have been found to limit the maximum achievable energy which can be
stored in a fusion plasma and may therefore be a problem for a future
reactor.
Concepts of stabilizing the plasma in order to handle these instabilities
include electron cyclotron current drive which can only be useful if it is
accurately adjusted to the needed quantity.
Therefore a thorough understanding of the island is necessary.

The time dependence of the magnetic island width $W$ is theoretically
described by the generalized Rutherford equation \cite{sau97}.
This first order nonlinear differential equation with respect to time contains in
our case three free parameters which have to be determined from measured data
since theoretical considerations can only provide estimates for these values.
They are assigned to three terms describing stabilizing and destabilizing
effects in the plasma, i.e.\ the bootstrap effect (with parameter $a_{BS}$), the
Glasser-Greene-Johnson effect ($a_{GGJ}$) and the polarization currents
($a_{pol}$).
We use a simplified form of the Rutherford equation which
comprises the relevant dependencies on the parameters
$\vec a^T = (a_{BS} , a_{GGJ} , a_{pol})$ only.
The full account of all physical constants and time dependent quantities
may be found in \cite{zoh97}.
\begin{equation}
\frac{dW(t)}{dt} = {\rm const} + a_{BS}  c_{BS}(t) \frac{W(t)}{W_{min}^2+W(t)^2}
                            - a_{GGJ} c_{GGJ}(t) \frac{1}{W(t)}
                            - a_{pol} c_{pol}(t) \frac{1}{W(t)^3}
\ .
\label{rutherford}
\end{equation}
$W_{min}$=1.8cm is the minimum width of an island.
The variables $c_{BS}$, $c_{GGJ}$ and $c_{pol}$ contain fundamental constants and
time dependent input quantities like plasma temperature or pressure.

\section{Generating the data}

The data is obtained from the so called Mirnov coils which are distributed
poloidally around the torus and measure any change of the poloidal
magnetic field.
The time variation $dm/dt$ of the magnetic flux $m(t)$ through the Mirnov
coil is proportional to the recorded signal.
We are interested in the time evolution of the amplitude $m$
of the integrated signal which can be connected to the magnetic island
width $W$ via
\begin{equation}
W(t) = \sqrt{\frac{m(t)-m_o}{b}}
\quad,
\label{model}
\end{equation}
where $m_o$ is the offset of the magnetic signal and $b$ a proportionality constant.
Additional information about the absolute size of the magnetic island for a certain
time comes from the electron cyclotron emission (ECE) diagnostic.
From this we know that $W_{ECE}=7$cm within a range of $\Delta_{ECE}=1$cm.
This information will be used later in setting up a prior.

\subsection{Extracting the data from the Mirnov signal}

\begin{figure}
\includegraphics[height=9cm]{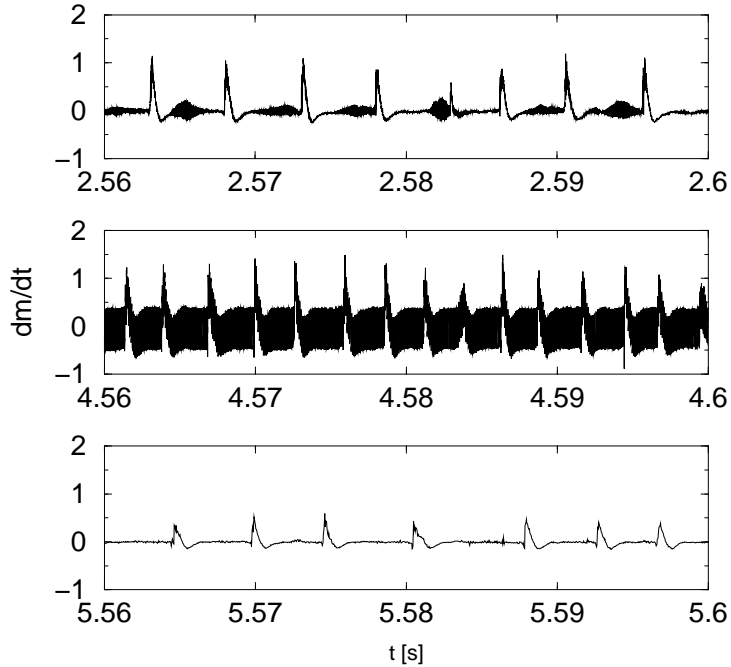}
\caption{
Signal from Mirnov coils for discharge \#12238 of the plasma device ASDEX
Upgrade for different times.
At $\sim 2.6$s the island has not formed yet, while at $\sim 4.6$s one can
see the induction of the magnetic island in the coil signal.
Finally at $\sim 5.6$s the island has disappeared.
The peaky structures have to be removed from the signal for further
process.
}
\label{figmirnov}
\end{figure}

\begin{figure}
\includegraphics[height=7cm]{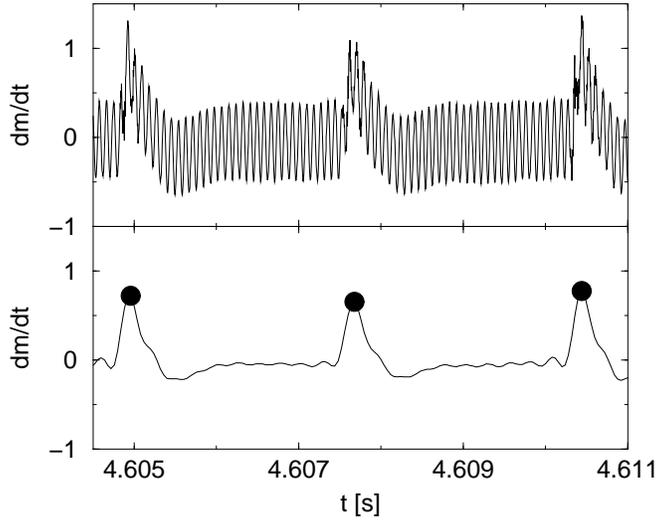}
\caption{Original signal (upper graph) and after Fourier transformation
back and forth where the higher frequency part was removed by filtering
(lower graph).
The peaky structures are easily identified (full circles).
}
\label{figfourier}
\end{figure}

The original signal from the Mirnov coils is shown in Fig.\
\ref{figmirnov}.
A closer look (upper graph in Fig.\ \ref{figfourier}) reveals two kinds of
structures:
On the one hand peaks at intervals of approximately 3-5ms which are due
to an edge plasma phenomena where energy and particles are expelled out of the
confined region, and
on the other hand the signal originating from the change of the
magnetic field which shows sinusoidal behavior (interval
approximately 0.08ms) with an amplitude
connected to the magnetic island width.
This is the information we want to extract.
First one has to identify the positions of the peaky structures.
Since the height of the peaks is not everywhere larger than the highest
amplitude of the sinusoidal signal we can not simply look for all points
which are higher than a certain level.
Fortunately the two structures live on two different scales in
frequency domain.
Therefore we Fourier transform (FFT) the complete data set and discard all the
higher frequencies which refer to the sinusoidal structure (see Fig.\
\ref{figfourier}).
Back transformation gives then a signal where peaks are easy to identify.
With the peak positions at hand we are set to go for the
amplitude of the sinusoidal structure in between two peaks.
Again Fourier transformation is applied where in addition we integrate over
time and are finally left with the magnetic signal $m$ shown in Figs.\
\ref{figseparation} and \ref{figresult}  .

\subsection{Finding the valid range of the model}

The Rutherford equation (\ref{rutherford}) describes the dynamics of
a magnetic island considering certain plasma physics effects.
However, at the onset of the mode the magnetic island is not stabilized and
subjected to fluctuations which are not covered by the model used.
We therefore have to identify the region in which the Rutherford equation is valid.
\begin{figure}
\includegraphics[height=9cm]{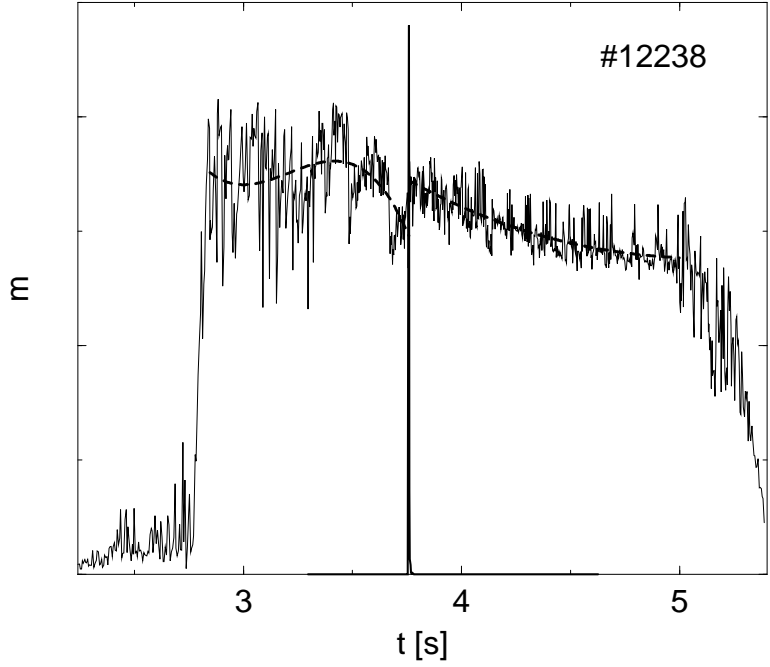}
\caption{
Amplitude of the magnetic signal of discharge \#12238.
After the onset of the magnetic island the signal has not saturated and
fluctuates (left side).
The analysis of the Rutherford equation can only be performed for the
stabilized part on the right side.
The full line is the posterior distribution for the change point $N_c$.
The dashed line is a fourth order polynomial in the respective regions.
}
\label{figseparation}
\end{figure}
Fig.\ \ref{figseparation} depicts the amplitude of the magnetic signal.
The left part differs from the right one where the island has stabilized
in amplitude and noise
and we have to look for the certain time incident $t_c=t(N_c)$
when the change of the behavior happens
-- a typical Bayesian change point problem.
Since we are out for the change point only
we describe the time variation of the amplitude of the
magnetic signal by low order polynomials
\begin{eqnarray}
m_i = \sum_{k=0}^K \alpha_k t_i^k \quad, \quad \forall \ i \le N_c \quad,
\\
m_i = \sum_{k=0}^K \beta_k  t_i^k \quad, \quad \forall \ i > N_c \quad.
\end{eqnarray}
For convenience we use matrix notation in the following, with matrix
elements $\{ \unbfm M \}_{k-1,i}=t_i^k$ and
$\vec \alpha^T = ( \alpha_0, ... ,\alpha_K)$.
The index $<$ ($>$) denotes time points before (after) the change point.
The data is corrupted by noise:
\begin{eqnarray}
\vec m_< = \unbfm{M}_< \vec \alpha + \varepsilon_< \quad,
\\
\vec m_> = \unbfm{M}_> \vec \beta + \varepsilon_> \quad.
\end{eqnarray}
Again we assume $\langle \varepsilon_{</>} \rangle = 0$ and 
$\langle \varepsilon_{</>} \rangle^2 = \sigma_{</>}^2$. 
Then the likelihood reads
\begin{eqnarray}
p ( \vec m | N_c , \vec \alpha , \vec \beta , \sigma_< , \sigma_> , I )
=
\frac{ 1 }{ ( 2 \pi \sigma_<^2 )^{N_c/2} }
\frac{ 1 }{ ( 2 \pi \sigma_>^2 )^{(N-N_c)/2} }
\quad
\qquad
\qquad
\qquad
\qquad
\phantom{a}
\nonumber\\
\cdot
\exp \left\{
- \frac{1}{2 \sigma_<^2}
( \vec m_< - \unbfm{M}_< \vec \alpha )^T
( \vec m_< - \unbfm{M}_< \vec \alpha )
- \frac{1}{2 \sigma_>^2}
( \vec m_> - \unbfm{M}_> \vec \beta )^T
( \vec m_> - \unbfm{M}_> \vec \beta )
\right\}
\ .
\label{seplikeli}
\end{eqnarray}
We need the posterior distribution for the change point.
With the help of Bayes theorem we get
\begin{equation}
p ( N_c | \vec m , I ) = \frac{ p ( N_c | I ) }{ p ( \vec m | I )}
p ( \vec m | N_c , I )
\quad.
\end{equation}
The nominator in the fraction is the prior distribution in absence of any
data which is a constant $p ( N_c | I ) = {\rm const}$
since no change point is preferred, but limits the possible values to $K+1 <
N_c < N-K-1$.
The marginal likelihood 
$p ( \vec m | N_c , I )$ is obtained by marginalizing over all parameters
in (\ref{seplikeli}).
\begin{equation}
p ( \vec m | N_c , I )
=
\int
{\rm d} \vec \alpha \
{\rm d} \vec \beta \
{\rm d} \sigma_< \
{\rm d} \sigma_> \
p ( \vec m | N_c , \vec \alpha , \vec \beta , \sigma_< , \sigma_> , I )
p ( \vec \alpha , \vec \beta | I )
p ( \sigma_< , \sigma_> | I )
\ \ .
\end{equation}
For the prior in $\vec \alpha$ and $\vec \beta$ we take a constant but use
Jeffreys prior for $p ( \sigma_< , \sigma_> | I ) = 1/(\sigma_< \sigma_>)$.
All integrations can be performed analytically and finally yield
\begin{eqnarray}
p ( N_c | \vec m , I )
& \propto &
\frac{1}{\sqrt{\det \unbfm{M}_<^T \unbfm{M}_<}}
\frac{
\Gamma \left( \frac{N_c - K}{2} \right)
}{
\left[
\vec m_<^T \vec m_< -
\vec m_<^T \unbfm{M}_< ( \unbfm{M}_<^T \unbfm{M}_<)^{-1} \unbfm{M}_<^T \vec m_< 
\right]^{(N_c-K)/2}
}
\quad
\qquad
\phantom{a}
\nonumber\\
& \cdot &
\frac{1}{\sqrt{\det \unbfm{M}_>^T \unbfm{M}_>}}
\frac{
\Gamma \left( \frac{N-N_c - K}{2} \right)
}{
\left[
\vec m_>^T \vec m_> -
\vec m_>^T \unbfm{M}_> ( \unbfm{M}_>^T \unbfm{M}_>)^{-1} \unbfm{M}_>^T \vec m_> 
\right]^{(N-N_c-K)/2}
}
\ .
\end{eqnarray}
The posterior change point distribution is shown in
Fig.\ \ref{figseparation} for a polynomial with order $K=4$.
We checked order three to five to get the same result.

\section{Likelihood and Prior}

The measured quantity in Eq.\ (\ref{model}) is the magnetic signal $m$ with
measurement uncertainty $\varepsilon$.
It is given in the form of a time series with $N$ successive events, where
we can write
\begin{equation}
m_i = m_o + b \cdot W_i(\vec a)^2 + \epsilon \quad, \qquad
i=1,...,N
\quad .
\label{error}
\end{equation}
Assuming that $\langle \epsilon \rangle=0$ and $\langle \epsilon^2 \rangle=\sigma^2$
we get by virtue of the principle of maximum entropy the likelihood
\cite{siv96}
\begin{equation}
p ( \vec m | m_o , b, \vec a , \sigma , I ) = \frac{1}{(2 \pi \sigma^2)^{N/2}}
\exp
\left\{ - \frac{1}{2 \sigma^2} \sum_{i=1}^N
         \left[ m_i - m_o - b W_i ( \vec a)^2 \right]^2
\right\}
\quad.
\label{likelihood}
\end{equation}
Next step is the assignment of prior distributions for
the conditional dependencies on $m_o$, $b$, the free parameters $\vec a$
and $\sigma$.
Due to the above mentioned treatment of Fourier transforming the data back
and forth we loose any information about the actual scatter originating
from the measurement process.
All we know is that a variance $\sigma$ exists and that it functions like a
scale parameter -- justified reasons for employing Jeffreys prior.
\begin{equation}
p(\sigma|I)=\frac{1}{\sigma}
\quad.
\label{priorsigma}
\end{equation}
Looking at the data before or after the formation of the magnetic island
provides an estimate $\tilde m_o$ and its uncertainty
$\sigma_{\tilde m_o}$ for the offset and leads to a Gaussian prior
distribution
\begin{equation}
p ( m_o | \tilde m_o , \sigma_{\tilde m_o} , I ) = 
\frac{1}{\sqrt{2 \pi} \sigma_{\tilde m_o}} 
\exp \left\{ - \frac12 \frac{(m_o - \tilde m_o)^2}{\sigma_{\tilde m_o}^2} \right\}
\quad.
\label{priorm}
\end{equation}
The ECE measurement is a constraint on the range of the proportionality constant
$b$.
Given a certain value for the offset $m_o$ 
we insert $W_{ECE}$ in Eq.\ (\ref{model})
which gives an estimate $b_{ECE}=(m_{ECE}-m_o)/W_{ECE}^2$.
The ECE measurement uncertainty $\Delta_{ECE}$ may be used in order to set
up an upper and lower limit:
$b_{up/low}=(m_{ECE}-m_o)/(W_{ECE} -/+ \Delta_{ECE})^2$
A convenient prior function given an estimate within boundaries is the
beta prior \cite{gcs95}.
However, it operates for values between 0 and 1 only, so we have to renormalize
$x(b)=(b-b_{low})/(b_{up}-b_{low})$:
\begin{equation}
p ( b | m_o , W_{ECE} , \Delta_{ECE} , m_{ECE} , I ) =
\frac{\Gamma(u) \Gamma(v)}{\Gamma(u+v)} \cdot
x(b)^{u-1} [1-x(b)]^{v-1}
\quad,
\label{priorb}
\end{equation}
where
\begin{equation}
u=\frac{(1-\mu_b) \mu_b - \sigma_b^2}{\sigma_b^2} \cdot \mu_b
\quad, \qquad
v= \frac{1-\mu_b}{\mu_b} \cdot u
\end{equation}
and
\begin{equation}
\mu_b=x(b_{ECE})
\quad , \qquad
\sigma_b = \frac{2}{b_{up}-b_{low}} \frac{b_{ECE}}{W_{ECE}} \Delta_{ECE}
\quad.
\end{equation}
From theoretical considerations we have some idea about the quantities of the
free parameters but unfortunately only for certain ideal configurations of the
confined plasma.
$a_{BS}=1.7$, $a_{GGJ}=6 \cdot 5/9$ and $a_{pol}=7$ are provided by
literature \cite{sau97,wil96}.
The maximum entropy principle gives us in this case an exponential function.
\begin{equation}
p ( \vec a | \vec a_o , I )= \prod_{j=1}^3 \frac{1}{a_{oj}}
\exp \left\{ - \frac{a_j}{a_{oj}} \right\}
\quad.
\label{priora}
\end{equation}
\section{Parameter Estimation}

We are out for the parameters of the Rutherford equation $\vec a$ together with
$m_o$ and $b$ from Eq.\ (\ref{model}).
In the expectation value for a component of
$\vec \theta^T = (m_o, b, a_{BS}, a_{GGJ}, a_{pol})$
we marginalize over all variables entering the likelihood Eq.\
(\ref{likelihood}):
\begin{equation}
\langle \theta_j \rangle =
\frac{
\int {\rm d} \theta_j \ \theta_j \ p ( \theta_j | \vec m , I )
}{
\int {\rm d} \theta_j \ p ( \theta_j | \vec m , I )
}
=
\frac{
\int {\rm d} \vec \theta \ \theta_j \int {\rm d} \sigma \
p ( \vec \theta , \sigma | \vec m , I )
}{
\int {\rm d} \vec \theta \ \int {\rm d} \sigma \
p ( \vec \theta , \sigma | \vec m , I )
}
=
\int {\rm d} \vec \theta \ \theta_j \ \rho (\vec \theta)
\quad.
\end{equation}
$\rho (\vec \theta)$ may be used as a sampling density in Markov chain
Monte Carlo (MCMC).
Invoking Bayes theorem
\begin{equation}
p ( \vec \theta , \sigma | \vec m , I ) =
\frac{
p ( \vec m | \vec \theta , \sigma , I ) p ( \vec \theta , \sigma | I )
}{
p ( \vec m | I )
}
\end{equation}
gives
\begin{equation}
\rho( \vec \theta ) =
\frac{
\int {\rm d} \sigma \
p ( \vec m | \vec \theta , \sigma , I ) p ( \vec \theta , \sigma | I )
}{
\int {\rm d} \vec \theta \ \theta_j \ \int {\rm d} \sigma \
p ( \vec m | \vec \theta , \sigma , I ) p ( \vec \theta , \sigma | I )
}
\quad.
\label{margrho}
\end{equation}
The full prior in Eq.\ (\ref{margrho}) disentangles into the functions given
in Eqn.\ (\ref{priorm},\ref{priorb},\ref{priora},\ref{priorsigma})
\begin{equation}
p ( \vec \theta , \sigma | I ) =
p ( m_o | \tilde m_o , \sigma_{\tilde m_o} , I )
p ( b | m_o , W_{ECE} , \Delta_{ECE} , m_{ECE} , I )
p ( \vec a | \vec a_o , I )
p ( \sigma | I)
\quad.
\end{equation}
The integration over $\sigma$ can be treated analytically and results in
\begin{equation}
\int {\rm d} \sigma \
p ( \vec m | \vec \theta , \sigma , I ) p ( \sigma | I)
\propto
\left\{ \sum_{i=1}^N
\left[ m_i - m_o - b W_i ( \vec a)^2 \right]^2
\right\}^{-\frac{N-1}{2}}
\quad.
\end{equation}
The final integrations over the parameters $\vec \theta$ are performed
numerically by employing MCMC, while the first order differential equation
(\ref{rutherford}) is solved applying second order Runge-Kutta method.

\section{Results}

The analysis is performed for discharge \#12238 of the plasma device
ASDEX Upgrade.
\begin{figure}
\includegraphics[height=9cm]{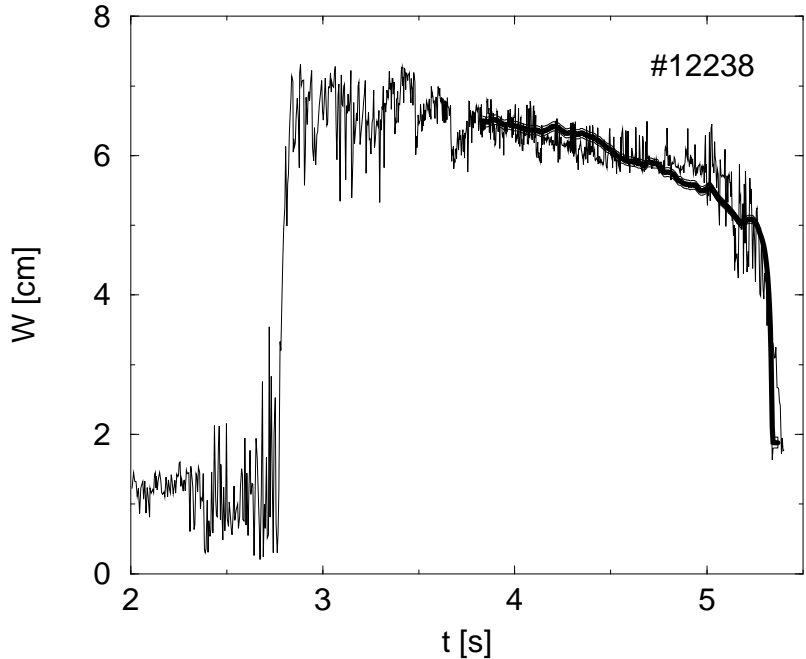}
\caption{Magnetic island width for discharge \#12238 of ASDEX Upgrade.}
\label{figresult}
\end{figure}
Fig.\ \ref{figresult} depicts the dynamics of the magnetic island width.
The thin line is the width obtained from the magnetic signal $m$ employing
Eq.\ \ref{model} with the expectation values of $m_o$ and $b$ from the
analysis (see table \ref{tabresult}).
\begin{table}[b]
\begin{tabular}{
r@{\hspace{.1cm}$\pm$\hspace{.1cm}}l
r@{\hspace{.1cm}$\pm$\hspace{.1cm}}l
r@{\hspace{.1cm}$\pm$\hspace{.1cm}}l
r@{\hspace{.1cm}$\pm$\hspace{.1cm}}l
r@{\hspace{.1cm}$\pm$\hspace{.1cm}}l}
\hline
   \multicolumn{2}{c}{$m_o[10^{-4}]$}
 & \multicolumn{2}{c}{$b$}
 & \multicolumn{2}{c}{$a_{BS}$}
 & \multicolumn{2}{c}{$a_{GGJ}$}
 & \multicolumn{2}{c}{$a_{pol}$}\\
\hline
 0.582 & 0.075 & 0.184 & 0.004 & 0.770 & 0.007 & 1.26 & 0.04 & 0.783 & 0.035\\
\hline
\end{tabular}
\caption{Expectation values with error margins for the magnetic offset
$m_o$, linear factor $b$ and the three parameters of
the Rutherford equation.}
\label{tabresult}
\end{table}
Only that time interval of the complete
signal is examined which comprises the island after is has stabilized
until the temperature
signal shows decoupling from the behavior of the collapsing island.
The comparison with the experimental data (thin line) gives a very good
agreement.
The accompanying parameters are given in table \ref{tabresult}.\\

\section{Summary}

Bayesian analysis was employed in order to identify the valid region in a data
set needed for further examinations.
The evolving data was used to determine free parameters in
the Rutherford equation, a first order nonlinear differential equation describing the
magnetic island dynamics in toroidally confined plasmas.


\begin{thebibliography}{5}

\bibitem{sau97}
Sauter, O., et~al., {\em Phys. Plasmas}, {\bf 4}, 1654 (1997).

\bibitem{zoh97}
Zohm, H., et~al., {\em Plasma Phys. Controlled Fusion}, {\bf 39}, B237
  (1997).

\bibitem{siv96}
Sivia, D.~S., {\em Data Analysis: A Bayesian Tutorial}, Clarendon Press,
  Oxford, 1996.

\bibitem{gcs95}
Gelman, A., Carlin, J., Stern, H., and Rubin, D., {\em Bayesian Data
  Analysis}, Chapman \& Hall, London, 1995.

\bibitem{wil96}
Wilson, H., et~al., {\em Phys. Plasmas}, {\bf 3}, 248 (1996).

\end{thebibliography}
\end{document}